\documentclass{statsoc}

\usepackage{natbib}

\usepackage{amssymb}
\usepackage{amsmath,amsfonts,verbatim,graphicx,float}
\usepackage{subfigure}
\usepackage{cite}

\newfont{\tensy}{cmsy10}

\title[Comments on the Unified approach \ldots]
{Comments on the Unified approach to the construction of Classical confidence intervals}

\author[W.~Wittek {\it et al.}]{Wolfgang Wittek, Hendrik Bartko, Nicola Galante and Thomas Schweizer}
\address{Max-Planck-Institut f\"ur Physik, D-80805 M\"unchen, Germany.}

\begin{document}
%
%





{\it Address for correspondence:} Wolfgang Wittek, 
Max-Planck-Institut f\"ur Physik, \\
\hspace*{3cm}D-80805 M\"unchen, Germany. E-mail: wolfgang.wittek@gmx.net

\begin{abstract}
The paper comments on properties of the so-called
"Unified approach to the construction of classical confidence intervals",
in which confidence intervals are computed in a Neyman construction
using the likelihood ratio as ordering quantity. In particular, two of the main
results of a paper by Feldman and Cousins (F\&C) are discussed.
It is shown that in the case of central intervals
the so-called flip-flopping problem, occuring in the specific scenario
where the experimenter decides to quote a standard upper limit or a confidence
interval depending on the measurement, is due to an expectation which
is not justified. The problem can be easily avoided by choosing
appropriate confidence levels for the standard upper limits and confidence
intervals.
In the F\&C paper "upper limit" is defined as the upper edge of a confidence interval,
whose lower edge coincides with the physical limit.
With this definition of upper limit (F\&C limit),
in an approach which uses the likelihood ratio as ordering quantity,
two-sided confidence intervals automatically change over to "upper limits"
as the signal becomes weaker (Unified approach). In the present paper
it is pointed out that this behaviour is not a special property of this
approach, because approaches with other ordering principles, like central intervals,
symmetric intervals or highest-probability intervals, exhibit the same
behaviour. The term "Unified approach" is therefore equally well justified
for these approaches.
The Unified approach is presented in the F\&C paper as a solution to the
flip-flopping problem. This might suggest that the F\&C limit is a
standard upper limit. Because the F\&C limit can be
easily misunderstood as standard upper limit, its coverage properties are
investigated.
It is shown that the coverage of the F\&C limit,
if it is interpreted as standard upper limit,
depends strongly on the parameter $\mu$ to be determined, with
values around and $\gtrsim (1+\alpha)/2$, where $\alpha$ is the confidence
level of the confidence belt. Differences between the F\&C
limit and a standard upper limit were already pointed out in the
F\&C paper. In order to exclude any misunderstanding, it is
proposed in the present paper to call the F\&C limit "upper edge of
the confidence interval", even if its lower edge coincides with the
physical limit.
\end{abstract}

%
%
%
\section{Introduction}
In 1998 the paper "Unified approach to the classical statistical analysis
of small signals" [\citet{FaC1998}] appeared, in which the classical approach
to the construction of confidence intervals is discussed in detail.
The paper has received great attention
and has stimulated the discussion about the calculation
of confidence intervals. Two of the main results of the paper can be stated as
follows:
\begin{itemize}
  \item By using the likelihood ratio as ordering quantity one obtains
confidence intervals which automatically change over from two-sided intervals
to upper limits
as the signal becomes statistically less significant
("Unified approach").
  \item This eliminates undercoverage caused by basing this choice (of
quoting an upper limit or a confidence interval) on the data
("flip-flopping").
\end{itemize}

These attractive features have induced many experimenters to follow the
approach proposed by Feldman and Cousins.
The purpose of the present paper is to
point out possible misunderstandings of the above statements and to give
alternative solutions to the flip-flopping problem.

Another result of the paper [\citet{FaC1998}] is, that an approach which
uses of the likelihood ratio as ordering quantity yields finite and
physical intervals, in cases where central intervals are empty or unphysical.
This subject is not discussed here.
%
%
%
\section{Neyman construction of confidence intervals and coverage}
\label{Section:Neyman}
%
%
%
To be specific the same example is considered as in [\citet{FaC1998}], a Poisson
process with background, where the mean background is known:
\begin{eqnarray}
P(n|\mu )\;=\;\dfrac{(\mu +b)^n}{n!}\times \exp[-(\mu +b)]
\label{eq:Pnmu}
\end{eqnarray}
\begin{tabular}{ll}
$n$           & is the measured number of signal plus background events \\
$\mu$         & is the average number of signal events \\
$b$           & is the average number of background events,
                $b$ is assumed to be exactly known \\
$P(n|\mu )$   & is the probability of measuring $n$, given $\mu$ and $b$
\end{tabular} \\

\noindent For the discussion in this Section and for later comparisons
with the results from [\citet{FaC1998}], the likelihood ratio
\begin{eqnarray}
R(n,\mu)\;=\;\dfrac{P(n|\mu )}{P(n|\mu_{best})}
\label{eq:likratio}
\end{eqnarray}
\noindent is assumed as ordering principle, when constructing the
acceptance region in
$n$ for an individual $\mu$. $\mu_{best}$ is the value of $\mu$ which maximizes
$P(n|\mu )$, at fixed $n$, where only physically allowed values of $\mu$ are
considered. For a given $\mu$, values of $n$ are added to the {\bf acceptance
region} $[n_1(\mu),\;n_2(\mu)]$ in decreasing order of the ratio $R(n,\mu)$, until
the sum $p\;=\;\sum_n P(n|\mu)$
agrees with the desired confidence level $\alpha$. Such an ordering
principle is naturally implied by the theory of likelihood ratio tests.

For a given measurement $n$,
the {\bf confidence interval} $I\;=\; [\mu_1,\;\mu_2]$ for $\mu$ is found by
including in $I$ all those $\mu$ which contain $n$ in their acceptance region.
The limits $\mu_1$ and $\mu_2$ obviously depend on $n$. The set of
confidence intervals $[\mu_1(n),\;\mu_2(n)]$ is called {\bf confidence belt}.

Upper limits of $\mu$ are determined in the same way as confidence intervals,
except that the ordering quanity is now $R(n,\mu)\;=\;n$. The resulting
"acceptance regions" are one-sided in this case: $[n_{low}(\mu), \infty]$.
The upper limit
$\mu_{up}(n)$ of $\mu$ at fixed $n$ is given by that $\mu$, for which the
lower edge $n_{low}(\mu)$ of the
acceptance region coincides with $n$. In general, upper limits of $\mu$ at a
certain confidence level differ from the upper ends of confidence intervals
at the same confidence level (see below).

Considering an ensemble of experiments with arbitrary but fixed $\mu$,
this fixed value of
$\mu$ will be contained in the confidence belt in a fraction $\alpha$ of all
experiments. This follows from the way the confidence intervals are
constructed: Given $n$, the confidence interval is the set of all those $\mu$
which
contain $n$ in their acceptance region. Confidence intervals with this property
 are said to have exact
coverage. Coverage is an important feature of Neyman-constructed confidence
intervals [\citet{Neyman1937}].

For upper limits the corresponding statement reads: Given an ensemble of
experiments with arbitrary but fixed $\mu$, this fixed value of $\mu$
will be less than
$\mu_{up}(n)$ in a fraction $\alpha$ of all experiments. In the following,
this definition of upper limit will be called "standard definition", and the
corresponding upper limit as "{\bf standard upper limit}".

It is evident that for both the confidence belt and for the standard upper limit
the specification of the confidence level is essential. For the confidence belt,
the ordering principle is relevant in addition. The ordering principle for
the standard upper limit is fixed.

In the present example small under or overcoverage occurs due to the
discreteness of $n$. In order to avoid this problem, in the calculations
presented below the factorial in the discrete Poisson function
$P(n|\mu )$ is replaced by Euler's Gamma function, which is then
normalized to 1. Euler's Gamma function, which is defined for all real
values of $n$, agrees with the discrete Poisson function at all integer
arguments $n$. The normalization factor differs from 1 by less than
0.1\% for $\mu\geq 3$, the maximum difference occuring at $\mu = 0$,
where it is less than 2 \%.

It should be stressed that it is not proposed to do this replacement in
practical applications. It is applied here only for the purpose of this paper,
to allow a discussion which is not affected by under or overcoverage due to
the discreteness of $n$.

In the general Neyman construction the ordering principle is not specified. In
the present paper, besides the ordering principle based on the likelihood ratio
(\ref{eq:likratio}) also the ordering principle based on central intervals is
considered.
%
%
%
\section{Discussion of Neyman-constructed confidence intervals and standard upper
limits}
\label{section:neyman}
%
%
%
In the previous Section the construction of confidence intervals and of
standard upper
limits is fully defined. In order to understand that this procedure of
defining confidence intervals and standard upper limits
ensures coverage, it is instructive to consider how
coverage would be tested by Monte Carlo simulations: One would generate
an ensemble of experiments with fixed $\mu_0$, throwing $n$ according to
eq.(\ref{eq:Pnmu}). An acceptance region $[n_1(\mu_0),\;n_2(\mu_0)]$
would be defined as explained in the previous Section. Although the confidence
belt is not yet completely defined by knowing the acceptance region for a single
$\mu_0$, one can already say that the confidence belt would contain $\mu_0$ in a
fraction $\alpha$ of all cases, because for those $n$ which lie in the
acceptance region of $\mu_0$ (which happens in a fraction $\alpha$ of all
cases)
$\mu_0$ would be added to the confidence interval  $[\mu_1(n),\;\mu_2(n)]$.
These statements are valid for all physically allowed values of $\mu$. One
can conclude that coverage is confirmed.

Thus, the way these confidence intervals and standard upper limits
are defined ensures exact coverage (disregarding effects which occur when the
problem involves discrete values). Obviously, the construction is independent
of the decision, whether to quote a standard upper limit, a confidence interval
or both.
Therefore, also coverage is guaranteed independent of such a decision.
The experimenter may thus decide to quote a standard upper limit, a confidence
interval or both, at equal or different confidence levels, without violating coverage.
Here, "coverage" is understood as explained in Section \ref{Section:Neyman}.
Coverage in a very specific scenario is discussed in Section
\ref{section:flip-flopping}.

\begin{table}
\caption{\label{tab:prob}Integrated probabilities $p_1$, $p$
and $p_2$ below, in an above the acceptance regions respectively, 
as functions of $\mu$.The confidence
level chosen is $\alpha$ = 90\%.}
\begin{tabular}{c|ccc}
 $ \;\;\;\;\mu\;\;\;\;$    &  $\;\;\;\;\;p_1\;\;\;\;$  &  $\;\;\;\;\;p\;\;\;\;$
&  $\;\;\;\;\;p_2\;\;\;\;$  \\
\hline
  0.0   &     0.000   &     0.9   &     0.100  \\
  0.5   &     0.000   &     0.9   &     0.100  \\
  1.0   &     0.016   &     0.9   &     0.084  \\
  1.5   &     0.035   &     0.9   &     0.065  \\
  2.0   &     0.046   &     0.9   &     0.054  \\
  2.5   &     0.053   &     0.9   &     0.047  \\
  3.0   &     0.056   &     0.9   &     0.044  \\
  3.5   &     0.057   &     0.9   &     0.043  \\
  4.0   &     0.057   &     0.9   &     0.042  \\
  4.5   &     0.057   &     0.9   &     0.043  \\
  5.0   &     0.056   &     0.9   &     0.044  \\
  5.5   &     0.056   &     0.9   &     0.044  \\
  6.0   &     0.056   &     0.9   &     0.044  \\
  6.5   &     0.056   &     0.9   &     0.044  \\
  7.0   &     0.056   &     0.9   &     0.044  \\
  7.5   &     0.056   &     0.9   &     0.044  \\
\hline
\end{tabular}
\end{table}
\begin{table}
\caption{\label{tab:limits}Lower and upper edge ($n_1$ and $n_2$)
of the 90\% c.l. acceptance region of $n$, 90\% c.l. lower limit and
95\% c.l. lower limit of $n$, as functions of $\mu$.}
\begin{tabular}{c|cc|cc}
  $\;\;\;\mu\;\;\;$    &  $\;\;\;\;n_1\;\;\;$  &  $\;\;\;\;n_2\;\;\;$
&  $\;\;\;n_{90\%}(\mu)\;$ &  $\;\;n_{95\%}(\mu)\;\;$ \\
\hline
  0.0   &     0.00   &     5.34   &     1.00  & 0.62 \\
  0.5   &     0.00   &     6.01   &     1.28  & 0.84 \\
  1.0   &     0.53   &     6.91   &     1.59  & 1.09 \\
  1.5   &     1.13   &     7.92   &     1.92  & 1.36 \\
  2.0   &     1.59   &     8.85   &     2.26  & 1.65 \\
  2.5   &     2.01   &     9.70   &     2.61  & 1.96 \\
  3.0   &     2.39   &    10.48   &     2.98  & 2.28 \\
  3.5   &     2.75   &    11.20   &     3.35  & 2.61 \\
  4.0   &     3.09   &    11.87   &     3.72  & 2.95 \\
  4.5   &     3.43   &    12.51   &     4.10  & 3.29 \\
  5.0   &     3.78   &    13.15   &     4.49  & 3.65 \\
  5.5   &     4.14   &    13.79   &     4.87  & 4.00 \\
  6.0   &     4.50   &    14.43   &     5.26  & 4.36 \\
  6.5   &     4.87   &    15.06   &     5.66  & 4.73 \\
  7.0   &     5.23   &    15.69   &     6.05  & 5.09 \\
  7.5   &     5.61   &    16.32   &     6.45  & 5.47 \\
\hline
\end{tabular}
\end{table}
\begin{table}
\caption{\label{tab:limitsofmu}Lower and upper edge ($\mu_1$ and $\mu_2$)
of the 90\% c.l. confidence interval of $\mu$, 90\% c.l. standard upper limit and
95\% c.l. standard upper limit of $\mu$, as functions of $n$.}
\begin{tabular}{c|cc|cc}
  $\;\;\;n\;\;\;$    &  $\;\;\;\;\mu_1\;\;\;$  &  $\;\;\;\;\mu_2\;\;\;$
&  $\;\;\;\mu_{90\%}(n)\;$ &  $\;\;\mu_{95\%}(n)\;\;$ \\
\hline
  0.0   &     0.00   &     0.61   &     0.00   &     0.00 \\
  0.5   &     0.00   &     0.98   &     0.00   &     0.00 \\
  1.0   &     0.00   &     1.38   &     0.02   &     0.86 \\
  1.5   &     0.00   &     1.90   &     0.87   &     1.75 \\
  2.0   &     0.00   &     2.50   &     1.64   &     2.58 \\
  2.5   &     0.00   &     3.16   &     2.36   &     3.35 \\
  3.0   &     0.00   &     3.87   &     3.04   &     4.08 \\
  3.5   &     0.00   &     4.59   &     3.71   &     4.81 \\
  4.0   &     0.00   &     5.30   &     4.39   &     5.50 \\
  4.5   &     0.00   &     6.01   &     5.04   &     6.21 \\
  5.0   &     0.00   &     6.69   &     5.67   &     6.89 \\
  5.5   &     0.11   &     7.36   &     6.31   &     7.56 \\
  6.0   &     0.49   &     8.04   &     6.94   &     8.21 \\
  6.5   &     0.81   &     8.69   &     7.56   &     8.89 \\
  7.0   &     1.04   &     9.34   &     8.19   &     9.54 \\
  7.5   &     1.29   &    10.01   &     8.81   &    10.19 \\
  8.0   &     1.54   &    10.64   &     9.41   &    10.84 \\
  8.5   &     1.81   &    11.29   &    10.03   &    11.47 \\
  9.0   &     2.08   &    11.93   &    10.64   &    12.11 \\
  9.5   &     2.37   &    12.56   &    11.24   &    12.74 \\
\hline
\end{tabular}
\end{table}

It should also be noted that coverage has to be obeyed at fixed $\mu_0$, and
for coverage it is irrelevant how far $\mu_0$ is away from the edges
$\mu_1(n)$ and $\mu_2(n)$ or whether $\mu_1(n)$ is given by the lowest
physically allowed value of $\mu$. This is characteristic for a Frequentist
approach. The discussion in Section \ref{section:unifiedapproach} will refer
to this feature.

\begin{figure}
\centering
\includegraphics[width=0.48\textwidth]{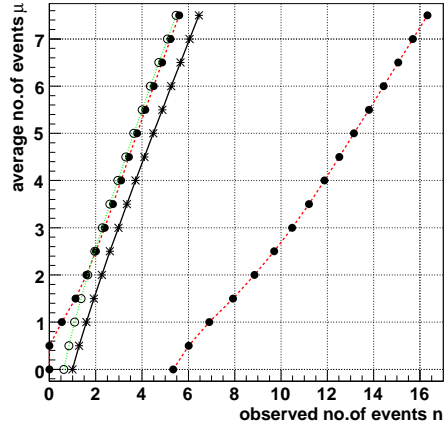}
\caption{\label{fig:FaC}90\% c.l. confidence belt (full circles), 90\%
c.l. standard upper limit (stars) and 95\% c.l. standard upper limit (open circles)
for the average number of signal events $\mu$
in the presence of a Poisson background with known mean $b=3.0$.}
\end{figure}

Assuming a confidence level of $\alpha$ = 90\%, setting $b$ equal to 3
and using the likelihood ratio as ordering quantity,
one obtains for the present example the confidence belt shown in
Fig.\ref{fig:FaC} (full circles).
Also shown in Fig.\ref{fig:FaC} are the 90\% c.l. standard
upper limit (stars) and the 95\% c.l. standard upper limit (open circles)
of $\mu$.

As can be seen, the 90\% c.l. standard upper limit and the 90\% c.l. confidence
belt in Fig.\ref{fig:FaC} are perfectly compatible with the corresponding limit
and confidence belt respectively in Figs. 5 and 7 of [\citet{FaC1998}].
Replacing the discrete Poisson distribution by the renormalized Euler's
function is therefore a valid procedure, for the purpose of this paper.

Table \ref{tab:prob} gives for each $\mu$ the integrated probabilities
below, in and above the acceptance region of $n$:
$p_1=\int_{n<n1}P(n|\mu ){\rm d}n$, $p=\int_{n1\leq n \leq n2}P(n|\mu ))
{\rm d}n$, $p_2=\int_{n>n2}P(n|\mu )){\rm d}n$. By construction, $p$ is
always equal to $\alpha$ = 90\%. As a result of the ordering principle chosen,
$p_1$ is generally different from $p_2$, $p_1$ being
less than $p_2$ for $\mu\;<\;2.4$ and greater than $p_2$ for
 $\mu\;>\;2.4$. 

In Table \ref{tab:limits} the lower and upper edge ($n_1$ and $n_2$)
of the 90\% c.l. acceptance region of $n$, the 90\% c.l. lower limit
$n_{90\%}(\mu)$ of $n$ and the 95\% c.l. lower limit $n_{95\%}(\mu)$
of $n$ are given as functions of $\mu$. The inverse functions
of $n_1(\mu)$, $n_2(\mu)$, $n_{90\%}(\mu)$ and $n_{95\%}(\mu)$
are the functions  $\mu_2(n)$, $\mu_1(n)$, $\mu_{90\%}(n)$ and
 $\mu_{95\%}(n)$ respectively, which define the confidence intervals
and standard upper limits of $\mu$ as functions of $n$. They are listed in Table
\ref{tab:limitsofmu}.

By definition, for a fixed $\mu_0$, $n$ lies within the acceptance region
$[n_1(\mu_0),\;n_2(\mu_0)]$ if and only if $\mu_0$ lies in the confidence
belt $[\mu_1(n),\;\mu_2(n)]$. This means that the probability that
$[\mu_1(n),\;\mu_2(n)]$ covers the value $\mu_0$ is equal to $p$ = $\alpha$.

Also by construction, $n>n_1(\mu )$ if and only if $\mu < \mu_2(n)$, implying
that the probability for $\mu < \mu_2(n)$ is equal to $p+p_2$. Thus
$[\mu_1(n),\;\mu_2(n)]$ is a confidence interval at the $\alpha$-confidence
level.
If $\mu_2(n)$ is interpreted as a standard upper limit, its coverage  $(p+p_2)$
depends strongly
on $\mu$, making this interpretation problematic.
The situation is different for central intervals, for
which $p_1= (1-\alpha )/2$, $p$ = $\alpha$ and
$p_2=(1-\alpha )/2$. In the case of central intervals, $\mu_2$ is
a standard upper limit at the confidence level $p+p_2\;=\;(1+\alpha)/2$, at all $\mu$.

%
%
%
\section{The flip-flopping problem}
\label{section:flip-flopping}
%
%
%
In this Section a specific scenario is considered,
in which the experimenter decides to quote a standard upper limit at the c.l. $\beta$ if $n$
is below and a confidence interval at the c.l. $\alpha$ if $n$ is above a certain
value $n_0$. It should be noted that this very specific scenario is different from the
scenario which is assumed when constructing confidence intervals or standard
upper limits
(see Section \ref{Section:Neyman}). In the latter scenario one talks independently
about coverage for a Neyman-constructed confidence interval or coverage for a
Neyman-constructed standard upper limit, and there is no condition involving the
measurement $n$. Therefore, the coverages for the two scenarios
are not expected to be identical. In addition, the coverage $\gamma$ for the specific
scenario depends on the choice of $\alpha$ and $\beta$. For central intervals all
coverages of interest can be directly given, without the need of complicated calculations.

In [\citet{FaC1998}] the special case $\alpha = \beta = 90$ \% is discussed for
an approach with central intervals. One obtains a coverage of
$\gamma =  \beta -p_2=\alpha -p_2= (3\alpha -1)/2=85$ \%, at low $\mu$.
The fact that $\gamma\neq \alpha$ is called
in [\citet{FaC1998}] violation of coverage, or flip-flopping problem. Talking of
"violation of coverage" means that
a coverage of $\gamma = \alpha$ is expected. As pointed out above, there is no
justification for that.

The actual problem is that $\gamma$ is different for different regions of $\mu$.
If $\gamma$ were the same for all $\mu$, one could quote a unique confidence level
($=\gamma$) for the specific scenario and there would be no problem.

The behaviour $\gamma\neq \alpha$ at low $\mu$ is due to the fact that
with the chosen
alternative between a standard upper limit (at a c.l. $\beta$=90 \%)
and a confidence interval (at the c.l. $\alpha$=90\%) the upper edge
$\mu_2(n)$ of the confidence belt, in the region $n<n_0$, is shifted. Knowing this,
it is easy to find a way of avoiding the flip-flopping problem: One has to define
alternatives which preserve the upper edge $\mu_2(n)$, at all $\mu$. In the case of
central intervals the alternatives are a $(1+\alpha)/2=$
95\% c.l. standard upper limit and a $\alpha=$
90\% c.l.confidence interval. With this choice the upper edge $\mu_2(n)$ of the
confidence belt is not changed, because for central intervals $\mu_2(n)$, if
interpreted as standard upper limit, has a c.l. of 95\%. Since the upper edge of
the confidence belt is preserved, coverage is fulfilled. Thus the flip-flopping problem
can be easily avoided by choosing appropriate confidence levels for the standard
upper limit and the confidence intervals.
\section{The approach by Feldman \& Cousins}
\label{section:unifiedapproach}
In the approach described in [\citet{FaC1998}] a confidence belt is determined according
to a Neyman construction, where the likelihood ratio is used as ordering quantity.
Coverage is therefore guaranteed for the confidence level $\alpha$ chosen:
If an experiment with fixed $\mu_0$ were repeated many times one would obtain
a set of confidence intervals $[\mu_1(n), \mu_2(n)]$, and the relation
$\mu_1(n)<\mu_0 < \mu_2(n)$ would be correct in a fraction $\alpha$ of all cases.
Because for small $n$ the lower edge
$\mu_1(n)$ coincides with the physical limit of $\mu$, the set of
relations (which define the confidence belt) can also be written as
\begin{eqnarray}
\qquad\qquad \mu_0\;<\;\mu_2(n)\qquad\qquad {\rm for}\;\; n\leq n_0 \qquad\qquad
                                             {\rm (a)} \nonumber\\
                                             \label{eq:FCuplim} \\
\qquad\qquad \mu_1(n)\;<\;\mu_0\;<\;\mu_2(n)\qquad\qquad {\rm for}\;\; n>n_0 \qquad\qquad
                                             {\rm (b)} \nonumber
\end{eqnarray}
where $n_0$ is the largest $n$ for which $\mu_1(n)$ coincides with the physical
limit. For $n\leq n_0$, $\mu_2(n)$ is called "upper limit" in  [\citet{FaC1998}], and
it will be called F\&C limit in the following: the F\&C limit is defined as the
upper edge of a confidence interval, whose lower edge coincides with the physical
limit of $\mu$.
Again, out of all relations (\ref{eq:FCuplim}) a fraction $\alpha$ of them would
be staisfied, and coverage would be fulfilled. However, one should not call the F\&C limit
an "$\alpha$ c.l. upper limit", because the relation  $\mu_0\;<\;\mu_2(n)$ is not
satisfied in a fraction $\alpha$ of all cases, and, moreover, this fraction depends on $\mu_0$.

In [\citet{FaC1998}] the presence of two kinds of intervals ((a) and (b))
is expressed as "This choice (of using the likelihood ratio as ordering quantity)
yields intervals which automatically change over from upper limits to two-sided
intervals as the signal becomes more statistically significant (Unified approach)".

This is correct, when the term "upper limit" is understood as the F\&C limit.
One should note, however, that this is not a special feature of
an approach which uses the likelihood ratio as ordering quantity.
Approaches with other ordering
principles, like central intervals, symmetric intervals or highest-probability
intervals, exhibit the same behaviour. The term "Unified approach" is therefore
equally well justified for these approaches.

The approach in [\citet{FaC1998}] is also presented as a solution to the
flip-flopping problem (see Section \ref{section:flip-flopping}). This could be
understood such that the approach in [\citet{FaC1998}] provides an alternative between a
standard upper limit and a confidence interval. However, this approach rather provides
an alternative between the F\&C limit and a confidence interval. The F\&C limit
can therefore be easily misunderstood as a standard upper limit.

As noted in Section \ref{section:neyman}, in a Frequentist approach
the value of the physical limit of $\mu$ is irrelevant, except that the
acceptance regions for $n$ can only be determined for physically allowed
values of $\mu$. The definition of "upper limit" as the upper edge of a
confidence interval whose lower edge coincides with the physical limit, is therefore
not a Frequentist type of definition and may lead to confusion.
In order to exclude any misunderstanding it is recommended in the present paper
to call the F\&C limit "upper edge of the confidence belt", even if the lower
edge coincides with the physical limit of $\mu$.
Differences between the F\&C limit and the standard upper limit were also pointed
out by the authors of [\citet{FaC1998}].

From this discussion it follows, that the coverage properties of the F\&C limit,
if interpreted as a standard upper limit, are of great interest.
From Table \ref{tab:prob} one can see that (for $\alpha=90$\%) this coverage $(p+p_2)$
varies between 94,4 \% and 100 \%. Compared to
the confidence level $\alpha$ of the confidence belt, the F\&C limit
$\mu_2$, if interpreted as standard upper limit,
is very conservative, with a coverage around $(1+\alpha)/2$=95\% or an
overcoverage around $(1+\alpha )/2-\alpha =(1-\alpha )/2$=5\%.

In the case of central intervals the upper edge of the confidence interval has
the nice property that it is a standard upper limit with a well-defined confidence
level, namely $(1+\alpha)/2$, at all $\mu$.

Coming back to the flip-flopping problem: In the approach by F\&C this problem
is avoided, because introducing the alternative between an F\&C limit and a
confidence interval doesn't change the original confidence belt,
for which coverage is fulfilled. This is analogous
to the approach with central intervals, where the original confidence belt
is preserved by choosing the alternative between a $(1+\alpha)/2$ c.l. standard upper limit
and a $\alpha$ c.l. confidence interval (see Section \ref{section:flip-flopping}).
%
%
%
\section{Summary and Conclusions}
%
%
%
The paper discusses two of the main results of [\citet{FaC1998}].
One of them concerns the so-called flip-flopping problem, which according to
[\citet{FaC1998}] exists for central intervals in the specific scenario, where
the experimenter decides to quote a standard upper limit or a confidence interval,
depending on the observed data. In the present paper it is shown that this
problem is due to an expectation which is not justified. The problem
can be easily avoided by choosing appropriate confidence levels
for the standard upper limit and the confidence belt.

Another result of [\citet{FaC1998}] concerns the Unified approach.
In [\citet{FaC1998}] the term "upper limit" denotes the upper edge of a
confidence interval whose lower edge coincides with the physical limit of $\mu$.
With this definition of upper limit, upper ends of confidence intervals
automatically change over to "upper limits" as the signal becomes weaker.
This transition from confidence
intervals to upper limits is not a special
property of the approach using the likelihood ratio as ordering quantity, because
the same transition takes place in approaches with different ordering priciples,
like central intervals, symmetric intervals or highest-probability intervals.
Thus the term "Unified approach" is equally well justified for these approaches.
The approach with central intervals has the additional advantage that the upper end
of the confidence interval
is a standard upper limit at a well defined confidence level, at all $\mu$.

The Unified approach is presented in the F\&C paper as a solution to the
flip-flopping problem, where the experimenter quotes alternatively a
standard upper limit or a confidence interval, depending on the measurement.
The F\&C limit can therefore easily be misunderstood as a standard
upper limit. For this reason the coverage properties of the F\&C limit are
investigated. It is shown that the coverage of the F\&C limit,
if it is interpreted as standard upper limit,
depends strongly on the parameter $\mu$ to be determined.
Compared to the confidence level $\alpha$ of the confidence belt,
the upper limit is very conservative, with a coverage around $(1+\alpha )/2$ or
an overcoverage around $(1+\alpha )/2-\alpha =(1-\alpha )/2$. More precisely,
for $\alpha\;=\;90$ \%, the coverage of the upper limit varies between
94,4 \% and 100 \%, and a unique confidence level cannot be assigned to it.
Differences between the F\&C
limit and a standard upper limit were already pointed out in
[\citet{FaC1998}]. In order to exclude any misunderstanding, it is
proposed in the present paper to call the F\&C limit "upper edge of
the confidence interval", even if its lower edge coincides with the
physical limit.

In this paper only one example was discussed in detail. The conclusions and
recommendations, however, are also valid for any other application, in which
confidence and upper limits are determined using the likelihood ratio
as ordering quantity. Though, the applications are restricted to the cases
with only one unknown ($\mu$) and no nuisance paramter, except if the
nuisance parameter is exactly known ($b$).

Finally it should be emphasized that the criticism expressed in this paper
does not refer to the Neyman construction of confidence intervals itself
or to the use of the likelihood ratio as ordering quantity.
\end{document}